\documentstyle[12pt]{article}

\def\be{\begin{equation}}
\def\ee{\end{equation}}
\def\bdi{\begin{displaymath}}
\def\edi{\end{displaymath}}
\def\br{\begin{eqnarray}}
\def\er{\end{eqnarray}}

\def\u2{\mid u\mid^2}

\def\ra{\rightarrow}

\def\RR{{\rm I\kern-.1567em R}}                              % Doppel R 
 \def\CC{{\rm C\kern-4.7pt                                    % Doppel C 
 \vrule height 7.7pt width 0.4pt depth -0.5pt \phantom {.}}} 
 \def\ZZ{{\sf Z\kern-4.5pt Z}}                                % Doppel Z 

\begin{document}

\begin{titlepage}
\vspace*{-2 cm}
\noindent

\vskip 3cm
\begin{center}
{\Large\bf Symmetries of generalized soliton models and
submodels on target space $S^2$. }
\vglue 1  true cm
C. Adam and  J. S\'anchez-Guill\'en
\vspace{1 cm}

\vspace{1 cm}
{\footnotesize Departamento de F\'\i sica de Part\'\i culas,\\
Facultad de F\'\i sica\\
Universidad de Santiago\\
and \\
Instituto Galego de Fisica de Altas Enerxias (IGFAE) \\
E-15706 Santiago de Compostela, Spain}

\medskip
\end{center}

\normalsize
\vskip 0.2cm

\begin{abstract}
Some physically relevant non-linear models with solitons, which have target
space $S^2$, are known to have submodels with infinitly many conservation
laws defined by the eikonal equation. 
Here we calculate all the symmetries of these models and their submodels
by the prolongation method. We find that for some models, like the Baby
Skyrme model, the submodels have additional symmetries, 
whereas for others, like the Faddeev--Niemi model, they do not.

\end{abstract}

\vskip 2cm

\end{titlepage}
\section{Introduction }
Non-linear field theories which support extended, soliton type solutions
are of importance in various fields of physics, ranging from 
elementary particle theory to condensed matter.
If the fields of the theory are required to approach a constant value at
spatial infinity in order to guarantee finite action or energy, then 
space time $\RR \times \RR^d$ is topologically equivalent to $\RR \times S^d$
(here $d$ is the dimension of space). Static, soliton-like solutions are then 
maps $S^d \to {\cal N}$ (where ${\cal N}$ is the target space) and may
sometimes
be characterized by a topological index. It frequently happens that the 
energy of a field configuration can be bounded from below by that
topological index, which then implies the existence of non-trivial
soliton solutions for a non-trivial topological index.  If the target
space is a sphere $S^n$ then the maps are characterized by the elements
of the corresponding homotopy group $\pi_d (S^n)$. Here we shall consider
the target space $S^2$ and the space dimensions
$d=2,3$ with homotopy groups $\pi_2 (S^2)=\ZZ$ and $\pi_3 (S^2) =\ZZ$,
respectively.  

Among theories with target space $S^2$, a well-known theory 
in $2+1$ dimensional space
time is the $CP(1)$ or Baby Skyrme model with Lagrangian density
\be \label{cp1}
{\cal L}_2 = \frac{\partial_\mu u \, \partial^\mu \bar u}{(1+ u\bar u)^2}
\ee
where $u$ is a complex field, 
%%$u: \, \RR \times \RR^2 \to \CC$, which is
%%related to a map ${\bf n}: \, \RR \times \RR^2 \to S^2$, ${\bf n}^2 =1$,
%%via stereographic projection
which parametrizes the the stereographic projection of the target $S^2$
${\bf n}: \, \RR \times \RR^d \to S^2$, ${\bf n}^2 =1$, via
\be
{\bf n} = \frac{1}{1+ u\bar u} \, ( u+\bar u , -i ( u-\bar u ) , 
u\bar u -1 ) \; ;
\qquad
u  = \frac{n_1 + i n_2}{1 - n_3}.
\label{stereo}
\ee
We use greek indices for space-time components, $\mu ,\nu =0,1,2$ or
$\mu ,\nu =0,1,2,3$ and latin indices for space components, $j,k = 1,2$
or $j,k = 1,2,3$, respectively.

The probably best-known theory with target space $S^2$ in $3+1$ dimensions
is the Faddeev--Niemi model (\cite{Fad}, \cite{FN1}) with Lagrangian density
\be \label{FN-L}
{\cal L}_{\rm FN} = {\cal L}_2 - \lambda {\cal L}_4
\ee
where $\lambda $ is a dimensionful coupling constant, ${\cal L}_2$ is like
in Eq. (\ref{cp1}) (but with $\mu =0,\ldots , 3$) and ${\cal L}_4$ is
\be
 {\cal L}_4 = \frac{(\partial^\mu u \, \partial_\mu \bar u)^2 - (\partial^\mu
u \, \partial_\mu u)(\partial^\nu \bar u \, 
\partial_\nu \bar u)}{(1+u\bar u)^4} .
\ee
The Faddeev--Niemi model 
is the $S^2$ restriction of Skyrme theory and so
circumvents Derrick's theorem, because it consists
of two terms such that their corresponding energies behave oppositely under
a scale transformation. The existence of (static) soliton solutions for the 
lowest Hopf indices has been confirmed by numerical calculations 
(\cite{GH} -- \cite{HiSa}).

Further models with solitons may be constructed from the two Lagrangian
densities ${\cal L}_2$ and ${\cal L}_4$ separately by choosing appropriate
(non-integer) powers of these Lagrangians such that the corresponding energies
are scale invariant.\footnote{Non-polynomial Lagrangian densities of this 
type were first introduced
by \cite{DDI} as possible effective chiral pion models.}
For ${\cal L}_4$ the appropriate choice is
\be
{\cal L}_{\rm AFZ} = -({\cal L}_4)^\frac{3}{4}
\ee
and for this model infinitely many analytic soliton solutions were found
by Aratyn, Ferreira and Zimerman (=AFZ) by using an ansatz with toroidal
coordinates (\cite{AFZ1}, \cite{AFZ2}). 
We shall, therefore, refer to this model as the AFZ model
in the sequel. The analysis of the AFZ model was carried further in 
(\cite{BF}), where, among other results, all the space-time and (geometric) 
target space symmetries of the AFZ model were determined. 

The appropriate choice for ${\cal L}_2$ is
\be \label{Ni-La}
{\cal L}_{\rm Ni}= ({\cal L}_2)^\frac{3}{2}.
\ee
This model has first been proposed by Nicole (\cite{Ni}), and it was shown
in the same paper that the simplest Hopf map with Hopf index 1 is a
soliton solution for this model. To the best of our knowledge, there are no 
more results on this model available in the literature. We shall refer to
this model as the Nicole (=Ni) model in the sequel.  

All four models (Baby Skyrme, Faddeev--Niemi, AFZ and Nicole) have the same
target space $S^2$ described by the variable $u$, 
therefore they have some common properties. For instance,
all Lagrangians are invariant under modular transformations
\be
u \; \ra \; \frac{au + b}{-\bar b u +\bar a} \, , \quad a\bar a + b\bar b =1.
\ee
This is a simple consequence of the fact
that all four Lagrangians are scalars when
expressed in terms of the vector ${\bf n}$ and are, therefore, invariant
under $SO(3)$ rotations of this vector. 

Furthermore, the same area-preserving diffeomorphisms on the target space $S^2$
can be defined for all models, but this does not imply
that they are symmetries for all four field theories. In fact, only the
AFZ model has the area-preserving diffeomorphisms as symmetries, which may be
understood from the fact that the Lagrangian density of the AFZ model is
just (a power of) the pullback of the area two-form\footnote{So the
equations of motion are quadratic in time} on $S^2$ under the map
$u$ (\cite{BF}, \cite{FR}). 
For the other three models the generators $Q^G$ of the area-preserving
diffeomorphisms (to be defined below) do not generate
symmetries and the corresponding Noether currents $J^G_\mu$ are not conserved. 
However, as it was realized within the generalization of the zero curvature
representation, \cite{AFSG},
there exist submodels for all three theories such that these 
currents are conserved. The submodels are defined by a further
condition (in addition to the equation of motion), which is the same for
all three models (up to dimensionality), namely the complex eikonal
equation
\be \label{eik}
\partial^\mu u\partial_\mu u =0.
\ee
For fields $u$ which obey the equation of motion of the Baby Skyrme,
Faddeev--Niemi or Nicole model and, in addition, the complex eikonal
equation (\ref{eik}), the currents $J^G_\mu$ (to be defined below for
each model) are conserved for an arbitrary real function $G$ which depends
on both $u$ and $\bar u$ (but not on derivatives thereof), 
therefore these submodels have 
infinitely many conserved charges.

At this point a word of caution is appropriate: the existence of the
infinitely many conserved charges does {\em not} imply that these submodels
have the area-preserving diffeomorphisms as symmetries. The crucial issue is
that the complex eikonal equation is not of the Euler--Lagrange type, i.e., 
it does not result from an action principle. Therefore, there is no direct
relation between symmetries and conservation laws, and the issue of
symmetry has to be investigated separately for the submodels.
In any case, 
the existence of infinitely many conserved charges is nevertheless
quite restrictive and might simplify the analysis of these models, which is why
we call them integrable, in analogy to the situation in lower dimensions.

It is the purpose of this paper to investigate the symmetries of the
Baby Skyrme, Faddeev--Niemi and Nicole model and, especially, of their 
integrable submodels. Specifically, we study the symmetries of the equations
of motion (and of the eikonal equation) for static, time-independent fields
$u$, for simplicity and as they provide the 
%%because these are the ones relevant for 
soliton solutions.\footnote{Time evolution of solutions, included in principle
in the approach of \cite{AFSG}, can be most interesting \cite{AS-G}.}
In Section 2 we briefly review the issue of area-preserving diffeomorphisms
and of their infinitesimal generators, because we need them in the sequel.
In Section 3 we study the symmetries of the static complex eikonal equation
both in 2 and 3 dimensions, which will be relevant for the integrable
submodels. In Section 4 we study the symmetries of the static Baby Skyrme
model and of its integrable submodel. In Section 5 we do the same for the 
Nicole model, and in Section 6 for the Faddeev--Niemi model. Section 7
contains our conclusions. For the symmetry aspects of the AFZ model we
refer to (\cite{BF}, \cite{FR}, \cite{AS-G}).

\section{Area-preserving diffeomorphisms on $S^2$}

An area-preserving diffeomorphism on target
space is a transformation $u\ra v(u,\bar u)$ such that the area form on
$S^2$ remains invariant (see also Refs. \cite{BF} and \cite{FR}),
\be
\Omega \equiv \frac{1}{2i}\frac{dud\bar u}{(1+u\bar u)^2} =
 \frac{1}{2i}\frac{dvd\bar v}{(1+v\bar v)^2}.
\ee
For infinitesimal transformations $v=u+\epsilon $ it is easy to see that
the condition of invariance of the area form leads to
\be
\epsilon_u +(\epsilon_u)^* = 2\frac{\bar u \epsilon +u\bar \epsilon}{1
+u\bar u} .
\ee
Here subscripts mean partial derivatives, $\epsilon_u \equiv \partial_u 
\epsilon$. Further, we use overbars for the complex conjugate of a
variable, but stars to denote the operation of complex conjugation, e.g.,
$(u)^* = \bar u$, $(f_u)^* =\bar f_{\bar u}$, etc.
Defining 
\be
\epsilon =(1+u\bar u)^2 f \, ,\quad f=F_{\bar u}
\ee
the above equation for $\epsilon$ simplifies to
\be \label{Feq}
\partial_u \partial_{\bar u}(F+\bar F)=0
\ee
which is solved by any purely imaginary function $F$ of $u$ and $\bar u$. 

[Remark: it seems that the most general solution of (\ref{Feq}) is any
$F$ such that $F+\bar F =g(u) +(g(u))^*$; however, such a $g$ which
depends on $u$ only
may always be reabsorbed by redefining $F \ra F+g$ without changing $f$ or
$\epsilon $, therefore an arbitrary imaginary $F$ is the most  
general solution.]

Introducing the real function $G$ via $F=iG$, the area-preserving 
diffeomorphisms are therefore generated by the vector fields
\be
v^G =i(1+u\bar u)^2 (G_{\bar u} \partial_u -G_u \partial_{\bar u})
\ee
which obey the Lie algebra
\be \label{ap-alg}
[v^{G_1},v^{G_2}] = v^{G_3} \, ,\quad G_3 = i (1+u\bar u)^2
(G_{1,\bar u} G_{2,u} - G_{1,u}G_{2,\bar u}) .
\ee
For field theories with the two-sphere as target space
the generators of area-preserving diffeomorphisms
may be constructed from the canonical momenta $\pi$, $\bar\pi$ of the fields 
$u$ and $\bar u$. They read
\be
Q^G =i \int d^d {\bf x} (1+u\bar u)^2 (\bar \pi G_u -\pi G_{\bar u})
\ee
and act on functions of $u,\bar u, \pi ,\bar\pi$ via the Poisson bracket, where
the fundamental Poisson bracket is (with $x^0 =y^0$)
\be
\{ u({\bf x}),\pi ({\bf y}) \} = \{ \bar u({\bf x}),\bar \pi ({\bf y}) \}   
=\delta^d ({\bf x} - {\bf y})
\ee
as usual. The generators $Q^{G_i}$ close under the Poisson bracket,
$\{Q^{G_1} ,Q^{G_2} \} =Q^{G_3}$ where $G_3$ is as in (\ref{ap-alg}).

Via the Noether charges (generators of area-preserving diffeomorphisms)
$Q^G$ the action of infinitesimal area-preserving diffeomorphisms is
defined for all four field theories given in Section 1.

\section{Symmetries of the eikonal equation}

We shall use the method of prolongations for all our symmetry calculations,
and we shall use the symmetry-generating vector fields in 
evolutionary form. That is to say, when a vector field
$v=\delta u\partial_u +\delta\bar u\partial_{\bar u}$ acting on the target
space variables is given, then the coefficients for the prolongation of
the vector field (giving the action on derivatives of the field) are defined
as total derivatives of the original coefficient. Concretely, our equations
are at most second order, therefore we need the second prolongation of $v$,
\be
{\rm pr}^{(2)}v =\delta u\partial_u +\delta\bar u\partial_{\bar u}
+\delta u^j \partial_{u_j} +\delta\bar u^j \partial_{\bar u_j} +
\delta u^{jk} \partial_{u_{jk}} +\delta\bar u^{jk} \partial_{\bar u_{jk}}
\ee
where $u_j \equiv \partial_{x^j} u$, etc. and 
the Einstein summation convention is understood.
As said, the coefficients for the prolongations are given by total derivatives,
\be
\delta u = \phi \quad \Rightarrow \quad \delta u^j = D_j \phi\, ,\quad  
\delta u^{jk} = D_j D_k \phi .
\ee
The explicit expressions for these total derivatives depend on which 
dependencies are assumed for the function $\phi$. Here we assume that
$\phi$ may depend on the independent and dependent variables as well as
on the first derivatives of the latter, that is
\be \label{v-pr0}
\delta u = \phi ({\bf x},u,\bar u,u_j ,\bar u_j )
\ee
\be
\delta u^j = \phi_j + \phi_u u_j + \phi_{\bar u} \bar u_j + \phi_{u_k}
u_{jk} + \phi_{\bar u_k} \bar u_{jk}
\ee
and $\delta u^{jk} = D_k \delta u^j$ where, e.g., $D_k (\phi_u u_j)$ is
\be \label{v-pr2}
D_k (\phi_u u_j) = \phi_{ku} u_j +\phi_{uu}u_j u_k +\phi_{u \bar u}
u_j \bar u_k +\phi_{u u_l} u_j u_{kl} + \phi_{u \bar u_l} u_j \bar u_{kl}
+ \phi_u u_{jk}
\ee
(we do not display the full expression because of its length), 
etc. For the details of the used formalism we refer to the book (\cite{Olv}).

Now, a symmetry of a PDE $F(u,u_j ,u_{jk},\ldots )=0$ (of $n$-th order, say)
is a solution of the equation ${\rm pr}^{(n)} v(F)=0$ which holds on-shell,
i.e., when the original PDE is used together with its prolongations
(PDEs that follow from $F=0$ by applying total derivatives). 
We allow for a dependence of $\phi$ on ${\bf x},u,\bar u,u_j ,\bar u_j$,
therefore the resulting solutions will contain the geometric target space
symmetries $\phi = f(u ,\bar u)$, the geometric base space symmetries
(``space-time symmetries'') which in evolutionary form
are given by $\phi =A^j({\bf x})u_k$ (with real $A^j$ depending on 
the base space variables only) as well as ``generalized symmetries'' 
(where $\phi$ depends on $u_j ,\bar u_j$ in a more general fashion).

After these preparatory remarks let us do the explicit calculation for
the static eikonal equation in $d=3$ dimensions first,
\be \label{st-eik}
u_j u_j =0\, , \quad j=1,2,3
\ee 
(which generalizes immediately to the cases $d> 3$ as we shall see).
The action of ${\rm pr}^{(1)} v$ leads to
\be
( \phi_j + \phi_u u_j + \phi_{\bar u} \bar u_j + \phi_{u_k}
u_{jk} + \phi_{\bar u_k} \bar u_{jk}) u_j =0
\ee
which, with the help of (\ref{st-eik}) and its first prolongation
\be
u_j u_{jk} =0
\ee
gives
\be \label{eik-sy-3d-g}
\phi_j u_j + \phi_{\bar u} u_j \bar u_j +\phi_{\bar u_k} \bar u_{jk} u_j=0.
\ee
First we observe that there are no conditions on the $u$ and $u_j$ 
dependence of $\phi$, therefore $\phi$ may be an arbitrary function of
$u$ and $u_j$. Next, we use that by assumption nothing may depend on 
$\bar u_{jk}$, therefore the third term must vanish separately,
$\phi_{\bar u_k} \bar u_{jk} u_j=0$. This requires either $\phi_{\bar u_j}
=0$ or $\phi_{\bar u_j} \sim \bar u_j \, \Rightarrow \, \phi = \phi
(\bar u_j \bar u_j)$. But this argument vanishes on-shell, therefore we
may conclude that $\phi_{\bar u_j} =0$ without loss of generality. This
implies, in turn, that the second term in (\ref{eik-sy-3d-g}) must vanish
separately, because nothing may depend on $\bar u_j$, and , therefore,
$\phi_{\bar u}=0$. We are thus left with
\be \label{eik-sy-3d}
\phi = \phi ({\bf x},u,u_k ) \, , \quad \phi_j u_j =0.
\ee
This does {\em not} imply that $\phi$ cannot depend on ${\bf x}$, but it does
imply that $\phi$ cannot contain a term which solely depends on ${\bf x}$,
i.e. a $\phi = f({\bf x}) + \ldots$ is not permitted. An allowed term,
which will lead to the base space symmetries, is $\phi = A^j ({\bf x})
u_j$ which has to obey
\be \label{e-3d-bs}
\phi_k u_k \equiv A^j{}_k u_j u_k =0
\ee
with the solutions
\be \label{e-3d-ct}
A^j = {\rm const.}  \qquad {\rm or} \qquad 
A^j{}_k = - A^k{}_j  \qquad {\rm or} \qquad
A^j{}_k \sim  \delta^j_k 
\ee
providing thereby the generators of the conformal group in $d=3$ Euclidean
space. So we find the conformal group as the base space symmetry group
and the transformations $\phi =f(u)$ for an arbitrary function $f$ for
the geometric target space symmetries. 

Obviously there are many more symmetries besides the geometric symmetries
(i.e.,  base space
 and geometric target space symmetries). Here we want
to describe them briefly, because they are
of some independent interest,
although we shall 
not need the corresponding results for our purposes.
First of all,
obviously nothing depended on the fact that 
$d=3$, therefore all the above results are equally true for higher
dimensions $d>3$ (with the conformal base space transformations in $d$
dimensions, of course). Secondly, in addition to the base space symmetries 
(\ref{e-3d-ct}) and the geometric target space symmetries $\phi =
f(u)$ there are many more symmetries present for the eikonal equation.
Among these are generalized symmetries $\phi =f(u_j)$ for arbitrary $f$, or
non-projectable base space symmetries $\phi =A^j ({\bf x},u)u_j$, i.e., 
infinitesimal conformal transformations on base space where the parameters
of these transformations are arbitrary functions of $u$ rather than
constants. Even the dependence on ${\bf x}$ is more general than just the
conformal transformations (\ref{e-3d-ct}). Concretely, although each
$\phi = f(A^j u_j)$ with $A^j$ an infinitesimal conformal
transformation is a symmetry,
this is not the most general allowed ${\bf x}$ dependence. For instance,
for a $\phi$
which is quadratic in the $u_j$, that is
\be 
\phi = B^{jk}({\bf x}) u_j u_k ,
\ee
the solution $B^{jk} = A^j A^k$ is not the
most general solution to (\ref{e-3d-bs}). The most general solution still
depends on a finite number of parameters, but more than the 10 of the
conformal transformations, providing thereby a kind of generalization
of the conformal group in the context of generalized symmetries.  
Analogous results hold for higher powers of the $u_j$, i.e., for $\phi$
of the form $\phi = B^{j_1 \ldots j_n} ({\bf x}) u_{j_1} \ldots u_{j_n}$. 

Next, we want to discuss the symmetries of the eikonal equation in 
$d=2$ dimensions. Here it is useful to introduce the complex coordinate
$z=x^1 +ix^2$, in terms of which the eikonal equation in $d=2$ reads
\be
u_z u_{\bar z} =0
\ee
and the first prolongation of the vector $v=\delta u\partial_u + {\rm c.c.}$,
where $\delta u $ is given by 
$\delta u = \phi (z,\bar z,u,\bar u,u_z ,u_{\bar z},\bar u_z ,
\bar u_{\bar z})$, which has the first prolongation
\be 
\delta u^z =D_z \phi =\phi_z + \phi_u u_z +\phi_{\bar u} \bar u_z +
\phi_{u_z} u_{zz} +\phi_{u_{\bar z}} u_{z\bar z} +
+\phi_{\bar u_z} \bar u_{zz} + \phi_{\bar u_{\bar z}} \bar u_{z\bar z}
\ee
and a similar expression for $\delta u^{\bar z}$. The action of 
${\rm pr}^{(1)}v$ leads to
\be
\delta u^z u_{\bar z} + \delta u^{\bar z} u_z =0
\ee
and, upon using the eikonal equation and its two first prolongations,
to
\be
u_{\bar z}\phi_z +u_z \phi_{\bar z} +\phi_{\bar u} (\bar u_z u_{\bar z}
+\bar u_{\bar z} u_z ) +\phi_{\bar u_z} (\bar u_{zz}u_{\bar z} +
\bar u_{z\bar z} u_z ) + \phi_{\bar u_{\bar z}} (\bar u_{z\bar z}u_{\bar z} +
\bar u_{\bar z\bar z} u_z ) =0.
\ee
This requires $\phi_{\bar u}= \phi_{\bar u_z}= \phi_{\bar u_{\bar z}}=0$ and
leads to
\be
\phi_z u_{\bar z} +\phi_{\bar z}u_z =0
\ee
which has the general solution
\be \label{2d-eiksol}
\phi =F[u,u_z g(z,u),u_{\bar z} h(\bar z,u)]
\ee
where $F[\cdot ,\cdot ,\cdot ]$, $g(\cdot ,\cdot )$ and $h(\cdot ,\cdot )$ 
are arbitrary functions of their arguments. Again, the geometric target
space transformations $\phi =f(u)$ and the base space transformations
$\phi =g(z)u_z$ (conformal transformations in $d=2$) may be found among
the symmetry transformations. 

\section{Symmetries of the Baby Skyrme model}

The equation of motion for the Baby Skyrme model with Lagrangian (\ref{cp1}) in
$d=2$ dimensions is
\be \label{cp1-eom}
(1+u\bar u)\partial^\mu \partial_\mu u -2\bar u (\partial^\mu u )
(\partial_\mu u ) =0
\ee
and the canonical momentum $\pi$ for the field $u$ is ($\dot u\equiv
\partial_0 u$)
\be
\pi = (1+u \bar u)^{-2} \dot{\bar u}
\ee
leading to the generators for area-preserving diffeomorphisms 
$Q^G$ on the target $S^2$ and corresponding Noether currents $J^G_\mu$  
\be
Q^G = i \int d^2 {\bf x} (\dot u G_u - \dot{\bar u} G_{\bar u})
\; , \quad J^G_\mu = i (u_\mu G_u - \bar u_\mu G_{\bar u}) .
\ee
Their divergence is, with the help of the equations of motion,
\be
\partial^\mu J^G_\mu = i u^\mu u_\mu 
(\frac{2\bar u}{1+u\bar u}G_u + G_{uu}) -
i \bar u^\mu \bar u_\mu  
(\frac{2 u}{1+u\bar u}G_{\bar u} + G_{\bar u\bar u})
\ee
and, as said, the currents $J^G_\mu$ 
are not conserved (for general $G$)
for the Baby Skyrme model (\ref{cp1-eom}) but are
conserved for the submodel defined by the eikonal equation as an
additional condition.

For static, solitonic fields the field equation for the Baby Skyrme model is
\be \label{s-cp1}
(1+u\bar u) u_{jj} -2\bar u u_j u_j =0
\ee
and for its submodel they are
\be
u_{jj} =0 \quad {\rm and} \quad u_j u_j =0
\ee
or ($z=x^1 +ix^2$)
\be \label{s-cp1-sm}
u_{z\bar z}=0 \quad {\rm and} \quad u_z u_{\bar z} =0.
\ee

[Remark: The eikonal equation $u_z u_{\bar z}=0$ is closely related to the
Cauchy--Riemann equations, but it is slightly more general, being the product
of a holomorphic and an anti-holomorphic derivative and, therefore, 
also non-linear. As a consequence, the
resulting symmetries are slightly different from the symmetries 
of the Cauchy--Riemann equations, as well.]

[Remark: as far as solitons (static solutions of finite energy) are concerned, 
there is not much difference between the Baby Skyrme model and its integrable 
submodel. Each soliton solution of the Baby Skyrme model is a rational
function of the variable $z$ only, $u=R(z)$ (or of $\bar z$ only), and is,
therefore, a solution of the Cauchy--Riemann equations (or of its
anti-holomorphic counterpart). As a consequence, it is 
also a solution of the submodel, as was observed, e.g. in
\cite{Wer}.]

Now we want to discuss the symmetries both of the full static equation
(\ref{s-cp1}) and of its submodel (\ref{s-cp1-sm}). For the full model
(\ref{s-cp1}) the result of a long but fairly straight-forward
calculation (using the prolongations (\ref{v-pr0})--(\ref{v-pr2})) is
that the symmetry group is a direct product of base space and 
geometric target space transformations, $\phi = M(u) + A^j ({\bf x}) u_j$, 
where $M$ just generates the modular transformations, 
\be \label{mod-gen}
M= c^a M^a \; , \quad M^1 =\frac{1}{2}(1-u^2) \; ,\quad M^2 =\frac{1}{2}
(1+u^2) \; ,\quad M^3 =iu
\ee
where the $c^a$ are three real constants, and $A^j$ are the 
infinitesimal conformal transformations in $d=2$, i.e., $A^1 +i A^2
=f(z)$ where $z=x^1 +ix^2$ and $f$ is an arbitrary function of its argument.
As this result is well-known and hardly surprising, we do not show the 
details of the calculation.    

For the symmetry calculations of the submodel we may use the result 
(\ref{2d-eiksol}) from the eikonal equation that 
$\phi =F[u,u_z g(z,u),u_{\bar z} h(\bar z,u)]$. Further we need the
second prolongation
\be
\delta u^{z\bar z}=
D_{\bar z} D_z \phi =D_{\bar z}(\phi_z +\phi_u u_z +\phi_{u_z} u_{zz}
+\phi_{u_{\bar z}} u_{z\bar z}) = \ldots 
\ee
where we do not display the lengthy expression explicitly. The condition
${\rm pr}^{(2)}v(u_{z\bar z})=0$ just means $\delta u^{z\bar z}=0$ which,
using equations (\ref{s-cp1-sm}) and their first prolongations, leads
to
\be
\phi_{z\bar z} +\phi_{zu}u_{\bar z} + \phi_{\bar z u}u_z +\phi_{zu_{\bar z}}
u_{\bar z\bar z} + \phi_{\bar z u_z}u_{zz} +\phi_{u_z u_{\bar z}}
u_{zz}u_{\bar z\bar z}=0
\ee
where now each coefficient has to vanish separately. The condition
$\phi_{u_z u_{\bar z}}=0$ implies that $\phi$ is a direct sum of two
types of functions,
\be
\phi = F^1 [u,u_z f(z,u)] + F^2 [u,u_{\bar z}g(\bar z,u)]
\ee
and we assume now that $\phi = F^1 [u,u_z f(z)]$ (the second case may be
treated analogously). All conditions are then fulfilled identically
except for $\phi_{zu}=0$ which leads to
\be
F^1{}_{ab}u_z f_z +F^1{}_{bb} u_z^2 f_u f_z +F^1{}_b u_z f_{zu}=0
\ee
(here $F^1 \equiv F^1 [a,b]$, $a\equiv u$, $b\equiv u_z f(z,u)$) and is 
solved either by
\be
f_z \equiv 0 \quad \Rightarrow \quad \phi = G(u,u_z)
\ee
or by
\be
f_u =0 \; \wedge \; F^1{}_a =0 \quad \Rightarrow \quad \phi =F[u_z f(z)].
\ee
Of course, the coordinate transformations $\phi =u_z f(z)$ and the
geometric target space transformations $\phi =G(u)$ are among the
symmetries we found. Furthermore, we see that the target space symmetries 
are not related to the area-preserving diffeomorphisms, which would
require $\phi =i(1+u\bar u)^2 G_{\bar u}$ for an arbitrary real 
$G(u,\bar u)$.

All in all, we find that the submodel has the generalized symmetries
given by
\be
\phi = F[u_z f(z)] + G(u, u_z) +\tilde F[u_{\bar z}g(\bar z)] +
\tilde G(u, u_{\bar z})
\ee
and, therefore, has certainly more symmetries than the full Baby
Skyrme model. 

[Remark: Solutions to the submodel $u_k u_k =0 \; \wedge \; u_{kk}=0$ solve,
in fact, a much larger class of models. Indeed, if we start with the
Lagrangian density
\be
{\cal L}= f(u,\bar u)\partial^\mu u \partial_\mu \bar u
\ee
for a general real function $f$ of its arguments, then the
resulting Euler--Lagrange equation is
\be
f\partial^\mu \partial_\mu u +f_u \partial^\mu u\partial_\mu u=0
\ee
and, in the static case, is solved by any solution of the submodel.
The same reasoning also shows that the pair of equations
$u_k u_k =0 \; \wedge \; u_{kk}=0$ cannot have non-pathological solutions 
in more
than two dimensions, because they would solve the above equation of motion
and, thereby, violate Derrick's theorem (here pathological means that
any formal solution of the equation pair must lead to an infinite energy,
whatever the function $f(u ,\bar u)$ in the Lagrangian is).] 

\section{Symmetries of the Nicole model}

The Nicole model (\ref{Ni-La}) leads to the equation of motion
\be
\frac{1}{2} (u^\lambda \bar u_\lambda )_\mu u^\mu +
u^\lambda \bar u_\lambda u^\mu{}_\mu - \frac{u^\lambda \bar u_\lambda}{
1+u\bar u}(u^\mu \bar u_\mu u +3 u^\mu u_\mu \bar u) =0
\ee
and to the canonical momentum
\be
\pi =\frac{3}{2} (1 + u\bar u)^{-3} (u^\mu \bar u_\mu )^\frac{1}{2} 
\dot{\bar u}.
\ee
The Noether currents for the area-preserving diffeomorphisms are
\be
J^G{}_\mu = i (1+u\bar u)^{-1}(u^\lambda \bar u_\lambda )^\frac{1}{2}
(u_\mu G_u -\bar u_\mu G_{\bar u})
\ee
and their divergence may be computed with the help of the equation of motion
to be
\be
\partial^\mu J^G{}_\mu = i \frac{
(u^\lambda \bar u_\lambda )^\frac{1}{2}}{1+u\bar u} \left(
u^\mu u_\mu 
(\frac{2\bar u}{1+u\bar u}G_u + G_{uu}) -
 \bar u^\mu \bar u_\mu  
(\frac{2 u}{1+u\bar u}G_{\bar u} + G_{\bar u\bar u}) \right) .
\ee
Again, the currents are 
not conserved for general $G$ for the full Nicole model,
but they are conserved for the submodel when $u$ obeys the eikonal equation,
as well. 

The static field equation for the full Nicole model is
\be \label{Ni-stat}
\frac{1}{2}(u_{jk}\bar u_j u_k + \bar u_{jk}u_j u_k) +u_j \bar u_j u_{kk}
-\frac{u_j \bar u_j}{1+u\bar u}(u_k \bar u_k u +3 u_k u_k \bar u)=0
\ee
and for the submodel the equations are
\be \label{Ni-stat-ei}
\frac{1}{2} \bar u_{jk}u_j u_k  + u_j \bar u_j u_{kk}
-\frac{(u_j \bar u_j)^2 u}{1+u\bar u} = 0 \quad \wedge \quad u_j u_j =0.
\ee

[Remark: as already mentioned, the known results on this model are rather 
scarce. The only known analytic solution of the static equation (\ref{Ni-stat})
is the simplest Hopf map \cite{Ni}
\be
u= \frac{x^1 +ix^2}{2x^3 -i(1-r^2)}
\ee
(here $r^2 \equiv x^j x^j$). However, it is known that the simplest Hopf map 
also solves the eikonal equation (\cite{eik}), and, consequently, the submodel
(\ref{Ni-stat-ei}) (\cite{Wer}). 
In spite of the scarce results we know, therefore, that the
solution space of the submodel is non-empty.]

Now we want to calculate the symmetries of the static equations both for
the full Nicole model and for its submodel. For the full model a long
calculation, similarly to the case of the Baby Skyrme model, shows 
that the symmetry group is again a direct product of base space and 
geometric target space transformations, $\phi = M(u) + A^j ({\bf x}) u_j$, 
where, again,  $M$ generates the modular transformations, see (\ref{mod-gen}),
 and $A^j$ are the generators of the
infinitesimal conformal transformations in $d=3$, see (\ref{e-3d-ct}). 

For the calculation of the symmetries of the submodel we want to briefly
sketch the most important steps. We know from the symmetries of the eikonal
equation that $\phi $ is of the form $\phi =\phi({\bf x},u ,u_j)$, see
(\ref{eik-sy-3d}).
The action of the second prolongation of the symmetry-generating vector field 
on the submodel equation leads to
\bdi
\frac{1}{2} \bar \phi^{jk} u_j u_k + \bar u_{jk}\phi^j u_k + \phi^{jj} 
u_k \bar u_k +u_{jj} (\phi^k \bar u_k +\bar\phi^k u_k) -
\edi
\be
-2u_j \bar u_j (\phi^k \bar u_k +\bar\phi^k u_k )\frac{u}{1+u\bar u} -
(u_j \bar u_j)^2 \frac{\phi -u^2 \bar \phi}{(1+u\bar u)^2} =0.
\ee
There are two terms in this expression which contain third derivatives of
$u$, namely the first term, which contains $\bar \phi_{\bar u_l}\bar u_{jkl}
u_j u_l$ and the third term, which contains $u_j \bar u_j \phi_{u_l}
u_{kkl}$. With the help of the first prolongation of the field equation
the first term may be re-expressed like $\bar\phi_{\bar u_l} (-u_j \bar u_j
u_{kkl} +\ldots )$, where the remainder contains only second derivatives.
Cancellation of the two terms with third derivatives 
now requires $\bar \phi_{\bar u_l} =\phi_{u_l}$
which implies $\phi = \psi ({\bf x},u) + A^k ({\bf x}) u_k$. From the
symmetry results of the eikonal equation we may further conclude that 
$\psi =\psi (u)$ and that the $A^k$ are just the generators of the
conformal transformations in $d=3$ dimensions. It remains to determine
$\psi$.  
The coefficients of the terms with second derivatives give no further 
conditions (i.e., they either cancel mutually or vanish identically for 
a $\phi$
of the above form), and the terms containing only first derivatives 
give just one further condition,
\be
(u_j \bar u_j )^2 \left( \frac{1}{2}\bar \psi_{\bar u\bar u} -\frac{u
\bar \psi_{\bar u}}{1+u\bar u} - \frac{\psi -u^2 \bar \psi}{(1+u\bar u)^2}
\right) =0
\ee
which is solved by the modular transformations, $\psi(u) =M(u)$.
Therefore, the submodel does not have more symmetries than the full
Nicole model, in spite of the infinitely many conserved currents of
the former. 

\section{Symmetries of the Faddeev--Niemi model}

The Faddeev--Niemi model (\ref{FN-L}) leads to the equation of motion
\be
(1+u\bar u) \partial^\mu \left( u_\mu -2\lambda (1+u\bar u)^{-2} 
(\bar u^\nu u_\nu u_\mu -u^\nu u_\nu \bar u_\mu )\right) -
2\bar u u^\mu u_\mu =0
\ee
and to the canonical momentum
\be
\pi = \frac{\dot{\bar u}}{(1+u\bar u)^2} -2\lambda \frac{u^\mu \bar u_\mu
\dot{\bar u} - \bar u^\mu \bar u_\mu \dot u}{(1+u\bar u)^4} . 
\ee
The Noether currents for the area-preserving diffeomorphisms are
\bdi
J^G{}_\mu = i \left( u_\mu -\frac{2\lambda}{(1+u\bar u)^2} (u^\nu \bar u_\nu 
u_\mu - u^\nu u_\nu \bar u_\mu )\right) G_u -
\edi
\be
- i \left( \bar u_\mu -\frac{2\lambda}{(1+u\bar u)^2} (u^\nu \bar u_\nu 
\bar u_\mu - \bar u^\nu \bar u_\nu  u_\mu )\right) G_{\bar u}
\ee
and their divergence may be computed with the help of the equation of motion
to be
\be
\partial^\mu J^G{}_\mu = i  \left(
u^\mu u_\mu 
(\frac{2\bar u}{1+u\bar u}G_u + G_{uu}) -
 \bar u^\mu \bar u_\mu  
(\frac{2 u}{1+u\bar u}G_{\bar u} + G_{\bar u\bar u}) \right) .
\ee
Here, again, the currents are 
not conserved for general $G$ for the full Faddeev--Niemi model,
but they are conserved for the submodel where $u$ obeys the eikonal equation. 

[Remark: Observe that the above divergence does not contain a term
proportional to $\lambda$. This shows that the second Lagrangian ${\cal L}_4$
alone (remember that ${\cal L}_{\rm FN} = {\cal L}_2 -\lambda {\cal L}_4$,
see Eq. (\ref{FN-L})) is invariant under area-preserving diffeomorphisms as
well as arbitrary powers of this Lagrangian, demonstrating the invariance
of the AFZ model.] 

The static field equations for the full Faddeev--Niemi model are
\bdi
(1+u\bar u)^3 u_{kk} -2 (1+u\bar u)^2 \bar u u_k u_k -4\lambda u \left(
(u_j \bar u_j)^2 -u_j u_j \bar u_k \bar u_k \right) +
\edi
\be
+ 2\lambda (1+u\bar u) \left( \bar u_{jk} u_j u_k - u_{jk} u_j \bar u_k 
+\bar u_j u_j u_{kk} -u_j u_j \bar u_{kk} \right) =0,
\ee
and for the submodel they are
\be
(1+u\bar u)^3 u_{kk} -4\lambda u (u_j \bar u_j )^2 + 2 \lambda (1+u\bar u )
(\bar u_{jk} u_j u_k + u_j \bar u_j u_{kk} )=0
\ee
together with the eikonal equation.

[Remark: results on static solutions of the Faddeev--Niemi model are
again quite scarce. Only numerical results on the simplest solitons with
low topological index exist (\cite{GH} -- \cite{BS2}). Besides, in this case 
- contrary to the Nicole model - it is
not known whether the solution space of the submodel is empty or non-empty.
Recently a class of exact solutions (both static and non-static) has been
constructed in \cite{HS}, but the resulting solutions are not ``simple''
(e.g., they do not have obvious symmetries), and it is not known
whether they correspond, in the static cases, to true minima of the
energy within a given topological sector, or whether they are critical
points of another type.]

As the symmetry calculations are quite lengthy, we just quote the results here.
It turns out that again the submodel does not have more symmetry than the
full Faddeev--Niemi model and that the symmetries are just the expected ones.
The symmetry-generating vector field $v=\phi \partial_u$ is given by 
$\phi = M(u) +A^k({\bf x}) u_k$, 
where $M$ again generates the modular transformations
(\ref{mod-gen}), whereas $A^k$ this time
has to obey
\be
A^j = {\rm const.}  \qquad {\rm or} \qquad
A^j{}_k = - A^k{}_j  
\ee
i.e., it generates translations and rotations (conformal transformations being
absent because the static Faddeev--Niemi model is not scale invariant 
and contains the dimensionful coupling constant $\lambda$).

\section{Conclusions}

We have thoroughly analysed the symmetries of submodels of some relevant
solitonic relativistic theories defined in two and three  dimensions on
target space  $S^2$, which have an infinite number of conserved currents. They
arose in various attempts to  apply  a generalization of the  zero
curvature representation to the $CP^1$ (Baby Skyrme), the Faddeev and Niemi
$S^2$ restriction  of Skyrme theory  and  proposals to overcome the Derrick
scaling  with one of the terms, specially the  quadratic one due to Nicole,
as the quartic AFZ case has been already extensively studied. All submodels are
parametrized by a complex field and defined by the eikonal equation
$(\partial u)^2 =0$.

The {\it prolongation} method may be not so well-known in this physical
context. We have therefore done some  effort to explain it, giving 
detailed expressions. Furthermore, we calculated 
the canonical momenta and Poisson structures,
which may be useful for future work on these theories. This should be the
case, as well, for the analysis of the eikonal equation and area preserving
diffeomorphisms. Specifically, for the eikonal equation we found a rather
large symmetry, which may be of independent interest.

The general result for all cases is that the  area-preserving
diffeomorphisms are not symmetries of any eikonal submodel. Also, the 
three-dimensional  submodels of Faddeev--Niemi and Nicole have no additional
symmetries compared to the full theories.

The Baby Skyrme model is special, 
as the  restriction does have an intriguing
additional symmetry. This can be important as there is not much difference
of the solutions of the full model and the restriction, at least for the
static case. Finally we remind that the method can be easily  extended to
include the time dependence. 

\begin{table}

\begin{tabular}{||r|c|l|c|c||}
\hline
 & $\infty $ many & geometric & generalized & solutions \\
 model & conserv. laws & symmetries & symmetries & known \\
\hline
Baby Skyrme & yes$^a$ & $C_2 \times SU(2) $ & no & yes \\
\hline
submodel & yes & $C_2 \times C_2 $ & yes & yes \\
\hline
Nicole model & no & $C_3 \times SU(2)$ & no & yes \\
\hline
submodel & yes & $C_3 \times SU(2)$ & no & yes \\
\hline
Faddeev--Niemi & no & $E_3 \times SU(2)$ & no & yes \\
\hline submodel & yes & $E_3 \times SU(2)$ & no & no \\
\hline 
\end{tabular} 

\caption{Symmetries and conservation laws of the three soliton models and
their submodels.  \newline
$C_d \; \ldots \; $ conformal group in $d$ dimensions. \newline
$E_d \; \ldots \; $ Euclidean group (translations and rotations) in $d$
dimensions. \newline
${}^a $due to the infinite-dimensional base space symmetries $C_2$. 
}

\end{table}

Finally, we want to present our results on the symmetries of the three
models in Table 1.\\ \\ \\
{\large\bf Acknowledgement:} \\
The authors thank L. Ferreira for helpful remarks on the paper.
This research was partly supported by MCyT(Spain) and FEDER
(FPA2002-01161), Incentivos from Xunta de Galicia and the EC network
"EUCLID". Further, CA acknowledges support from the 
Austrian START award project FWF-Y-137-TEC  
and from the  FWF project P161 05 NO 5 of N.J. Mauser.


\begin{thebibliography}{99}
\bibitem{Fad}
L.D. Faddeev, in ``40 Years in Mathematical Physics'', World Scientific, 
Singapore 1995.
\bibitem{FN1}
L.D. Faddeev, A.J. Niemi,
Nature 387 (1997) 58; hep-th/9610193.
%%\bibitem{FN2}
%%L.D. Faddeev, A.J. Niemi, hep-th/9705176.
\bibitem{GH}
J. Gladikowski, M. Hellmund,
Phys. Rev. D56 (1997) 5194; hep-th/9609035.
\bibitem{BS1}
R.A. Battye, P. Sutcliffe,
Proc. Roy. Soc. Lond. A455 (1999) 4305, hep-th/9811077.
\bibitem{BS2}
R.A. Battye, P. Sutcliffe, Phys. Rev. Lett. 81 (1998) 4798.
\bibitem{HiSa}
J. Hietarinta, P. Salo, Phys. Rev. D62 (2000) 081701.
\bibitem{DDI}
S. Deser, M.J. Duff, C.J. Isham,
Nucl.  Phys. B114 (1976) 29.
\bibitem{AFZ1} 
H. Aratyn, L.A. Ferreira and A. Zimerman, 
Phys. Lett. B456 (1990) 162.
\bibitem{AFZ2} 
H. Aratyn, L.A. Ferreira and A. Zimerman, 
Phys. Rev. Lett. 83 (1999) 1723.
\bibitem{BF} 
O. Babelon and L.A. Ferreira, 
JHEP 0211 (2002) 020.
\bibitem{Ni} D.A. Nicole, 
J. Phys. G4 (1978) 1363.
\bibitem{FR}
L.A. Ferreira, A.V. Razumov, 
Lett. Math. Phys. 55 (2001) 143; hep-th/0012176.
\bibitem{AS-G}
C. Adam, J. Sanchez-Guillen, 
J. Math. Phys. 44 (2003) 5243; hep-th/0302189.
\bibitem{AFSG}
O. Alvarez, L.A. Ferreira, and J. S\'anchez-Guill\'en, 
Nucl. Phys. B529 (1998) 689.
\bibitem{Olv}
P.J. Olver, ``Applications of Lie Groups to Differential Equations'',
Springer Verlag, New York, 1993.
\bibitem{eik}
C. Adam, J. Math. Phys. 45 (2004) 4017; math-ph/0312031.
\bibitem{Wer}
A. Wereszczynski, hep-th/0410148.
\bibitem{HS}
M. Hirayama, C-G. Shi,
Phys. Rev. D69 (2004) 045001; hep-th/0310042.
\end{thebibliography}
\end{document}